\documentclass{nature-tight}
\usepackage{graphicx}
\usepackage{amssymb} 

\newcounter{biblio}

\newcommand{\lisevi}{\ensuremath{\mbox{\ion{$^7$Li}{1}}}}
\newcommand{\lisixi}{\ensuremath{\mbox{\ion{$^6$Li}{1}}}}
\newcommand{\lisev}{\ensuremath{\mbox{$^7$Li}}}
\newcommand{\lisix}{\ensuremath{\mbox{$^6$Li}}}
\newcommand{\liratio}{\ensuremath{\mbox{\lisix} / \mbox{\lisev}}}
\newcommand{\LiI}{\ensuremath{\mbox{\ion{Li}{1}}}}
\newcommand{\litofe}{\ensuremath{[{\rm \mbox{\lisev} / Fe}]}}

\newcommand{\KI}{\ensuremath{\mbox{\ion{K}{1}}}}
\newcommand{\NaI}{\ion{Na}{1}}
\newcommand{\SI}{\ion{S}{1}}
\newcommand{\CaII}{\ion{Ca}{2}}
\newcommand{\FeI}{\ion{Fe}{1}}
\newcommand{\HI}{\ion{H}{1}}
\newcommand{\hi}{\ion{H}{1}}
\newcommand{\htwo}{H$_2$}
\newcommand{\bvalues}{$b$-values}
\newcommand{\bvalue}{$b$-value}
\newcommand{\fvalues}{$f$-values}
\newcommand{\lya}{Ly$\alpha$}
\newcommand{\kms}{\ensuremath{{\rm km~s^{-1}}}}

\newcommand{\nav}{$N_a(v)$}

\newcommand{\vlsr}{\ensuremath{v_{\rm LSR}}}
\newcommand{\SNR}{\ensuremath{{\rm SNR}}}
\newcommand{\abund}[1]{\ensuremath{A({\rm #1})}}
\newcommand{\fuse}{{\em FUSE}}
\newcommand{\hst}{{\em HST}}
\newcommand{\iue}{{\em IUE}}
\newcommand{\ga}{\gtrsim}
\newcommand{\la}{\lesssim}
\title{Observation of interstellar lithium in the
  low-metallicity Small Magellanic Cloud}

\author{J. Christopher Howk$^{1}$, Nicolas Lehner$^1$, Brian
  D. Fields$^{2,3}$ \& Grant J. Mathews$^1$}

\begin{document}

\maketitle

\begin{affiliations}
{\scriptsize
\item Department of Physics, Center for Astrophysics, University of
  Notre Dame, Notre Dame, IN 46556, USA
\item Department of Astronomy, University of Illinois at
  Urbana-Champaign, Urbana, IL 61801, USA 
\item Department of Physics, University of Illinois at
  Urbana-Champaign, Urbana, IL 61801, USA
}
\end{affiliations}

\begin{abstract}

  The primordial abundances of light elements produced in the standard
  theory of Big Bang nucleosynthesis (BBN) depend only on the cosmic
  ratio of baryons to photons, a quantity inferred from observations
  of the microwave background.\cite{Dunkley2009} The
  predicted\cite{Steigman2007,Cyburt2008,Fields2011} primordial
  \lisev\ abundance is four times that measured in the atmospheres of
  Galactic halo stars.\cite{Spite1982, Sbordone2010,Melendez2010} This
  discrepancy could be caused by modification of surface lithium
  abundances during the stars' lifetimes\cite{Korn2006} or by physics
  beyond the Standard Model that impacts early
  nucleosynthesis.\cite{Jedamzik2004,Pospelov2010} The lithium
  abundance of low-metallicity gas provides an alternative constraint
  on the primordial abundance and cosmic evolution of
  lithium\cite{Prodanovic2004} that is not susceptible to the {\it in
    situ} modifications that may affect stellar atmospheres.  Here we
  report observations of interstellar \lisev\ in the low-metallicity
  gas of the Small Magellanic Cloud, a nearby galaxy with one quarter
  of the Sun's metallicity.  The present-day \lisev\ abundance in the
  Small Magellanic Cloud is nearly equal to the BBN predictions,
  severely constraining the amount of possible subsequent enrichment
  of the gas by stellar and cosmic ray nucleosynthesis.  Our
  measurements can be reconciled with standard BBN with an extremely
  fine-tuned depletion of stellar Li with metallicity.  They are also
  consistent with non-standard BBN.

\end{abstract}



\begin{figure}
\begin{center}
\includegraphics[angle=0, width=3.3in]{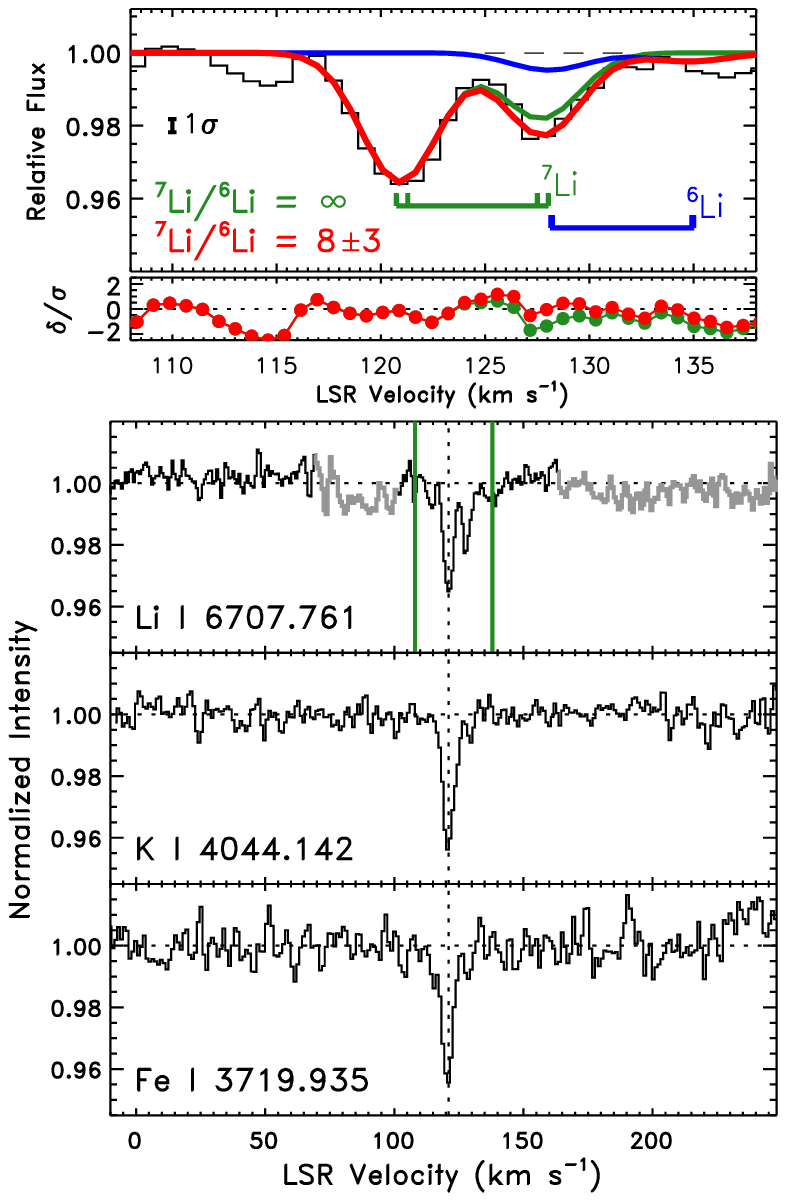}
\end{center}
\spacing{1.0}
\vspace{-0.75 cm}
\caption{\small {\bf Interstellar absorption by several neutral
    species seen toward Sk~143.}  Normalized interstellar absorption
  profiles from UVES plotted versus Local Standard of Rest (LSR)
  velocity and profile fit of the \LiI\ absorption.  The
empirically-determined \SNR\ is $\sim275$ per pixel (5 pixels per
resolution element) for the \LiI\ observations.  The full set of
optical and ultraviolet absorption profiles seen toward this star and
column densities measured from these are given in the Supplementary
Information.  The lower three panels show the profiles of \LiI, \KI,
and \FeI; the SMC cloud bearing \LiI\ at $\vlsr\approx +121$ \kms\ is
marked with the dashed line.  The grey regions near \LiI\ are possibly
contaminated by diffuse interstellar bands or residual fringing, which
may extend into the region containing Li absorption.  The effects on
the \lisevi\ columns are within the quoted uncertainties.  The
\LiI\ absorption is composed of (hyper)fine structure components of
both \lisevi\ and \lisixi\ (shown respectively by the green and blue
ticks in the top panel).  The strong line along of \lisevi\ is
detected with $\sim16\sigma$ significance in the ISM of the SMC.  A
model fit to the \LiI\ absorption complex is shown in the top panel
(see Supplementary Information), with the fit residuals shown
immediately below (normalized to the local error array).  The free
parameters for the fit are the polynomial coefficients for the stellar
continuum, the central velocity, Doppler parameter ($b$-value), and
column densities of \lisevi\ and \lisixi\ for the interstellar cloud.
The red curve shows the best fit joint fit including both \lisevi\ and
\lisixi, shown in green and blue, respectively.  The best fit isotopic
ratio is $N(\lisixi)/N(\lisevi) = 0.13 \pm 0.05$ (68\% c.l.),
consistent with the presence of \lisix\ along the sight line, although
below the $3\sigma$ detection threshold.  \label{fig:stack}}
\end{figure}

We obtained high resolution spectra ($R\approx70,000$) of the star
Sk~143 (AzV 456), an O9.5 Ib star in the Small Magellanic Cloud (SMC),
using UVES\cite{Dekker2000} on the 8.2-m VLT (observational details
are given in the Supplementary Information).  The sight line to this
star was chosen for observation because it shows significant
absorption from neutral atoms and molecules\cite{Cox2007, Welty2006,
  Cartledge2005} and a weak interstellar radiation
field,\cite{Welty2006} all of which favor the presence of neutral
lithium, \LiI.  \LiI\ absorption is clearly detected along this sight line (Figure
\ref{fig:stack}).

%
%
%
%

The derivation of the total Li/H abundance in the interstellar medium
(ISM) requires large corrections for ionization, since $ N({\rm Li})
\approx N(\mbox{\ion{Li}{2}}) \gg N(\mbox{\ion{Li}{1}})$, and for the
incorporation of Li into interstellar dust grains.\cite{Steigman1996}
Our first approach to these corrections uses observations of adjacent
ionization states of other metals, in this case Ca and Fe, to estimate
the amount of unseen gas phase lithium.  Assuming ionization balance
and only atomic processes, the ratio
$N(\mbox{\ion{Li}{2}})/N(\mbox{\ion{Li}{1}}) \propto
N(\mbox{\ion{Ca}{2}})/N(\mbox{\ion{Ca}{1}})$ or $\propto
N(\mbox{\ion{Fe}{2}})/N(\mbox{\ion{Fe}{1}})$, where the constant of
proportionality involves the ratios of ionization rates and
recombination coefficients for the elements in
question.\cite{Steigman1996, Welty2003} The ratio of \lisevi\ to total
hydrogen in the SMC is $\log [N(\mbox{\lisevi})/N({\rm H})] =
-11.17\pm0.04$ (all uncertainties are 1$\sigma$ unless noted), where
$N({\rm H}) \equiv N(\mbox{\HI}) + 2 N(\mbox{\htwo})$.  Applying
ionization corrections derived from Ca and Fe yields logarithmic
abundances $\abund{\lisev} \equiv \log [N(\lisev)/N({\rm H})] + 12 =
2.79\pm0.11$ and $3.01\pm0.12$.  These calculations do not include
more complicated (and uncertain) effects such as grain-assisted
recombination,\cite{Welty2003, Weingartner2001} nor do they correct
for dust depletion.

%

Our second approach uses the observation\cite{Steigman1996} that
$N(\mbox{\lisevi})/N(\mbox{\KI})$ along sight lines through the Milky
Way is nearly constant (with new determinations giving consistent
results\cite{Knauth2003, Welty2001}). When a differential ionization
correction is applied, \lisev /K in the Milky Way ISM is consistent
with the solar system ratio.  Thus, \lisev\ and K appear to have very
similar ionization and dust depletion behaviors, and \lisevi
/\KI\ gives a good measure of the total (gas+dust phase) \lisev /
K.\cite{Steigman1996, Knauth2003, Welty2001} We measure $\log
[N(\mbox{\lisevi})/N(\mbox{\KI})] = -2.27\pm0.03$ in the SMC, in
agreement with the Galactic relationship.\cite{Knauth2003,Welty2001}
Applying an ionization correction of $+0.54\pm0.08$
dex\cite{Steigman1996,Welty2003} gives $\log
[N(\mbox{\lisev})/N(\mbox{K})] = -1.78\pm0.09$.  With the solar system
ratio $\log (\mbox{\lisev}/\mbox{K})_\odot = -1.82\pm0.05$ derived
from meteorites,\cite{Asplund2009} we find $[\mbox{\lisev/K}]_{\rm
  SMC} \equiv \log [N(\mbox{\lisev})/N(\mbox{K})] - \log
(\mbox{\lisev}/\mbox{K})_\odot = +0.04\pm0.10$. The ratio of
\lisev\ to metal nuclei in the SMC is consistent with that found in
the solar system and the Milky Way ISM\cite{Steigman1996}: $(\lisev
/{\rm K})_{\rm SMC} \approx (\lisev /{\rm K})_\odot$.

Although the ionization and depletion characteristics of \ion{S}{1}
are not as well tied to those of \LiI,\cite{Welty2003} a similar
approach using \ion{S}{1} yields [\lisev /S]$_{\rm SMC} =
-0.26\pm0.11$.  The sub-solar ratio is consistent with a modest (0.3
dex) depletion of Li and K onto dust in the ISM\cite{Knauth2003}
relative to S.

We estimate \abund{\lisev} by scaling \lisev /K to Li/H:
$A(\lisev)_{\rm SMC} = A(\lisev)_\odot + [{\rm Fe/H}]_{\rm SMC} +
[{\rm K/Fe}]_{\rm SMC} + [{\rm \lisev/K}]_{\rm SMC}$.  We adopt $[{\rm
    ^7Li / K}]_{\rm SMC}$ from above, the meteoritic
$\abund{\lisev}_\odot = 3.23\pm0.05$,\cite{Asplund2009} with a mean
present-day SMC metallicity $[{\rm Fe/H}]_{\rm SMC} = -0.59\pm0.06$
and an SMC K/Fe abundance $[{\rm K/Fe}]_{\rm SMC} \equiv +0.00\pm0.10$
(the last two discussed in the Supplementary Information).  This
yields $\abund{\lisev}_{\rm SMC} = 2.68\pm0.16$.  Similarly scaling
the \lisev/S result gives $2.38\pm0.17$.



\begin{figure}
\begin{center}
  \includegraphics[angle=0, width=15.cm]{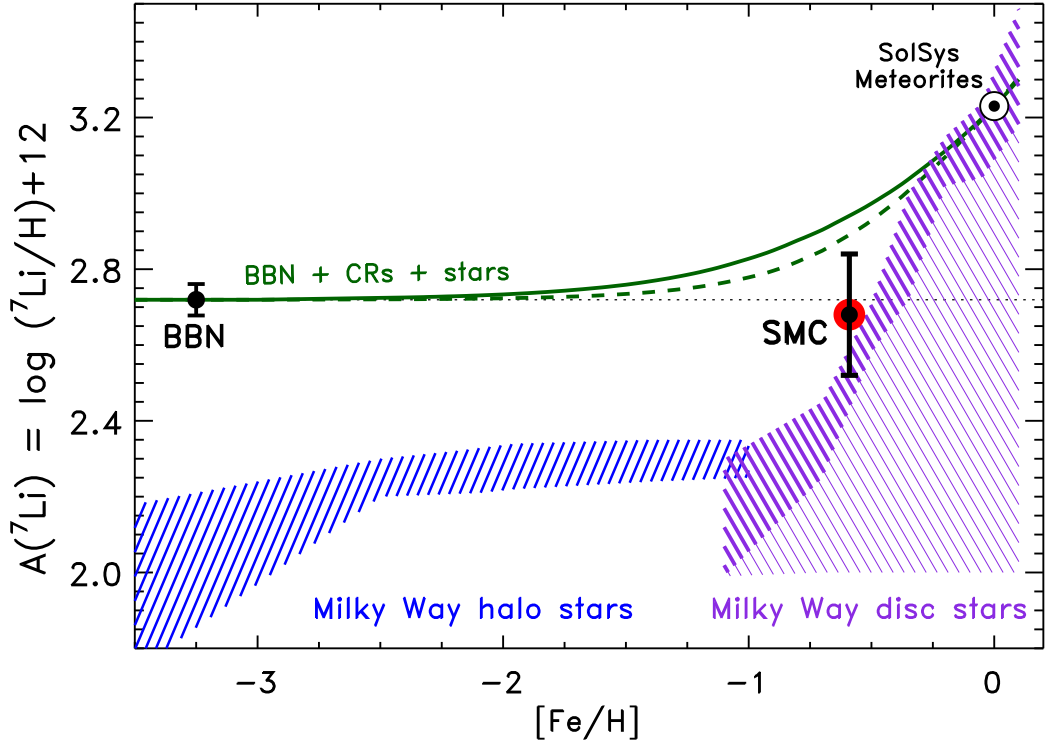}
\end{center}
\spacing{1.0} 
\vspace{-0.5 cm}
\caption{\small {\bf Estimates of the lithium abundance in the SMC
    interstellar medium and in several different environments.}  Our
  best estimate for interstellar gas+dust phase abundance of
  \abund{\lisev} in the SMC is shown as the red circle with black core
  derived from the \lisevi /\KI\ ratio.  The present day metallicity
  of the SMC from early-type stars is $[{\rm Fe/H}] = -0.59\pm0.06$.
  (All uncertainties are $1\sigma$.)  The point marked BBN and dotted
  horizontal line show the primordial abundance predicted by standard
  BBN.\cite{Cyburt2008} The green curves show recent
  models\cite{Prantzos2012} for post-BBN \lisev\ nucleosynthesis due
  to cosmic rays (CRs) and stars.  By adjusting the yields from
  low-mass stars, the models are forced to match the solar system
  meteoritic abundance\cite{Asplund2009} (see Supplementary
  Information).  The solid and dashed lines correspond to models A and
  B\cite{Prantzos2012} which respectively include or not a presumed
  contribution to \lisev\ from core-collapse supernovae.  The blue
  hatched area shows the range of abundances derived for Population II
  stars in the Galactic halo,\cite{Sbordone2010} with the ``Spite
  plateau'' in this sample at $\abund{\lisev}_{\rm Pop II} \approx
  2.10\pm0.10$.\cite{Sbordone2010} The violet hatched region shows the
  range of measurements seen in Galactic thin disk stars, where the
  thicker lines denote the six most Li-rich stars in a series of eight
  metallicity bins.\cite{Lambert2004} The selection of thin disk stars
  includes objects over a range of masses and temperatures, including
  stars that are expected to have destroyed a fair fraction of their
  Li.  Thus, the upper envelope of the distribution represents the
  best estimate of the intrinsic ISM Li abundance at the epoch of
  formation for those stars, and the thicker dashed lines for the thin
  disk sample are most appropriate for comparison with the SMC value.
  The most Li-rich stars in the Milky Way thin disc\cite{Lambert2004}
  within 0.1 dex of the SMC metallicity give $\abund{\lisev}_{\rm MW}
  = 2.54 \pm 0.05$, consistent with our estimate $\abund{\lisev}_{SMC}
  = 2.68\pm0.16$.  
  \label{fig:liabundance}}
\end{figure}

Most previous observational constraints on the primordial Li abundance
have relied on measurements of atmospheric abundances in
low-metallicity Galactic stars.  Our first detection of interstellar
lithium beyond the Milky Way opens a new window on the lithium
problem.  While there are significant uncertainties associated with
ionization and dust effects, as demonstrated by the significant spread
in $\abund{\lisev}_{\rm SMC}$ values, these are largely independent of
the uncertainties that might affect stellar measurements of the
primordial lithium abundance.  Our recommended absolute abundance is
$\abund{\lisev}_{\rm SMC} = 2.68\pm0.16$, or $(\lisev/{\rm H})_{\rm
  SMC} = (4.8\pm1.8)\times 10^{-10}$, derived from \lisev /K.  This is
compared to stellar \lisev\ abundances\cite{Sbordone2010,Lambert2004}
at different metallicities in Figure \ref{fig:liabundance}.  Our best
estimate overlaps the prediction from standard BBN using the baryonic
density deduced from the 5-year WMAP data,\cite{Dunkley2009}
$\abund{\lisev} = 2.72\pm0.06$ (95\% c.l.),\cite{Cyburt2008} although
this leaves little room for the post-BBN chemical
evolution,\cite{Prantzos2012,Romano2003} i.e., the contribution of
freshly-synthesized Li to the ISM by stellar and cosmic ray
nucleosynthesis (see representative models\cite{Prantzos2012} in
Figure \ref{fig:liabundance}).  Our estimate of $\abund{\lisev}_{\rm
  SMC}$ is also consistent with the upper envelope of Li abundances in
Milky Way thin disc stars (Figure
\ref{fig:liabundance}).\cite{Lambert2004}

\begin{figure}
\begin{center}
  \includegraphics[angle=0, width=15.cm]{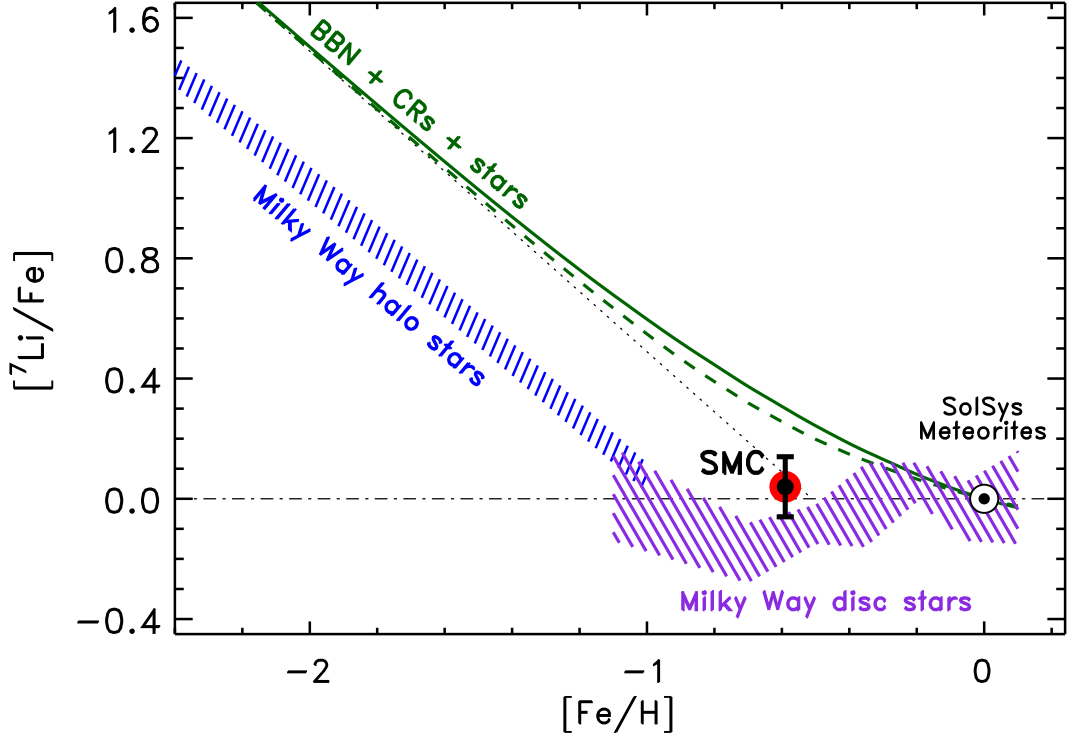}
\end{center}
\spacing{1.0} 
\vspace{-0.5 cm}
\caption{\small {\bf Estimates of Li/Fe in the SMC interstellar medium
    and in several different environments.} The SMC value is derived
  from the \lisevi /\KI\ ratio.  At low metallicities ($[{\rm
      Fe/H}]\la-1$), stellar
  measurements\cite{Sbordone2010} trace the build-up of Fe
  with a constant Li abundance along the Spite plateau.  At higher
  metallicities, disc star abundances\cite{Lambert2004} show a
  turnover to roughly constant \litofe\ at values consistent with the
  solar system/meteoritic value\cite{Asplund2009} (shown as the
  dash-dotted line).  Our SMC estimate is consistent with the solar
  system and disc star abundances in this region of relatively
  constant \lisev /Fe abundances, with $\litofe_{\rm SMC} =
  +0.04\pm0.14$ for $[{\rm K/Fe}]_{\rm SMC}=0.0\pm0.10$ (Supplemental
  Information).  The most Li-rich disc stars within 0.1 dex of the SMC
  metallicity have a mean $\langle [{\rm \lisev /Fe}]\rangle =
  -0.13\pm0.05$.  (All uncertainties are $1\sigma$.) The green curves
  show the chemical evolution models\cite{Prantzos2012} as in Figure
  \ref{fig:liabundance}, while the dotted line shows the behavior of
  \litofe\ for the standard BBN primordial abundance with no
  subsequent evolution of \lisev.  The relative uniformity of the
  stellar \lisev/Fe abundances at $[{\rm Fe/H}]\ga-1$ could be caused
  by a delicate balance of Li and Fe production and
  metallicity-dependent Li astration (not ruled out given the changes
  in mean age and mass potentially present in the
  sample\cite{Lambert2004}).  However, the agreement of the
  \litofe\ ratio seen in these old stars (ages $>$4
  Gyr\cite{Lambert2004}) and in the present-day interstellar medium of
  the SMC suggests little change in the stellar abundances for
  metallicities $[\rm {Fe/H}]\ga-0.6$ through the solar metallicity.
  To bring the stellar and SMC interstellar abundances into agreement
  with standard BBN predictions requires a delayed injection of
  significant \lisev\ from stellar production mechanisms as well as
  vigorous depletion of stellar surface \lisev\ abundances at
  metallicities just below that of the SMC.
  \label{fig:li2metal}}
\end{figure}

However, given the uncertainties in scaling to $\abund{\lisev}_{\rm
  SMC}$, the stronger result is our measurement $[{\rm ^7Li / K}]_{\rm
  SMC}= +0.04\pm0.10$.  We compare $[{\rm ^7Li / K}]_{\rm SMC}$ with
measurements\cite{Asplund2009,Sbordone2010,Lambert2004} of
\litofe\ and chemical evolution models\cite{Prantzos2012} in Figure
\ref{fig:li2metal}.
The stars show a rapid decrease in [\lisev /Fe] with increasing
metallicity until $[{\rm Fe/H}]\approx -1$, at which point the Li
abundance increases roughly in lockstep with Fe such that disc stars
have a nearly-constant \litofe\ ratio similar to that found in the
solar system.  Our measurement of the present day \lisev -to-metal
ratio in the SMC is in agreement with the nearly constant values found
in the atmospheres of Milky Way disc stars ($-1 \la [{\rm Fe/H}] \la
0$), most of which formed $>$4 Gyr ago, with the solar system, and the
modern-day Milky Way ISM.\cite{Steigman1996}

Both the thin disc stars and our SMC measurements are below standard
BBN predictions with reasonable assumptions about post-BBN production,
although it is often assumed these stars have had significant
depletion of their surface Li abundance.\cite{Prantzos2012} Taken at
face value, the consistency of our SMC measurement with the
\litofe\ for those stars calls this assumption into question.  While
the models in Figures \ref{fig:liabundance} and \ref{fig:li2metal} are
imprecise given the uncertain Li yields from stellar sources, they
illustrate the tension between standard BBN predictions and our
measurements if there is any post-BBN Li production.  This tension can
be relieved if a metallicity-dependent depletion of Li in stellar
atmospheres is fine tuned in such a way that it is very strong below
$[{\rm Fe/H}] \approx [{\rm Fe/H}]_{\rm SMC} = -0.6$ (to create the
Spite plateau and avoid overproducing Li in the SMC ISM) and
negligible at or above the SMC metallicity, conspiring to create a
constant \litofe\ ratio above $[{\rm Fe/H}] \approx -1$.
Alternatively, non-standard BBN scenarios can be invoked to allow for
a lower primordial Li abundance.\cite{Iocco2009,Fields2011}

If non-standard Li production occurs in the BBN epoch, many such
models predict excess \lisix\ compared with standard BBN. The only
known source of post-Big Bang \lisix\ is production via cosmic ray
interactions with ISM particles.  Excess \lisix\ at the metallicity of
the SMC would support non-standard production mechanisms, either in
the BBN epoch\cite{Pospelov2010} or through the interaction of
pre-galactic cosmic rays with intergalactic helium.\cite{Suzuki2002}
Measurements of \lisix\ in stellar atmospheres are extremely difficult
since the stellar line broadening is well in excess of the isotope
shift.  However, the \lisevi\ doublet is well-separated in our data
due to the very low broadening in the cool ISM probed by
\LiI\ absorption.  Our best fit to the SMC \LiI\ absorption gives
$(\liratio)_{\rm SMC} = 0.13 \pm 0.05$ (see Supplementary Information
and Figure \ref{fig:stack}), giving a formal limit to the isotopic
ratio in the SMC of $(\liratio)_{\rm SMC} < 0.28$ ($3\sigma$).  With
higher signal-to-noise and resolution it should be possible to lower
the limits for the interstellar isotope ratio in the SMC to a point
that they provide constraints on non-standard BBN models.  This
approach has the advantage that ionization and dust depletion effects
are not important for comparing the two isotopes of
Li,\cite{Kawanomoto2009} making \liratio\ a powerful diagnostic of
nucleosynthesis and non-standard evolution of Li abundances.


\vfill

\hrule

\hrule


\begin{addendum}
\item[Supplementary Information] is linked to the online version of
  the paper at www.nature.com/nature.

\item[Acknowledgements] We thank the European Southern Observatory for
  granting us time for this project as part of proposal 382.B-0556.
  We also thank A. Fox and H. Sana for helpful discussions about the
  UVES data and A. Korn, P. Molaro, T. Prodanovic, D. Romano, and
  D. Welty with helpful input on the project that improved the paper.

\item[Author Contributions] All authors participated in the
  interpretation and commented on the manuscript.  J.C.H. led the
  project and was responsible for the text of the paper.
\item[Author Information] Correspondence and requests for materials
  should be addressed to J.C.H.~(email: jhowk@nd.edu).
\end{addendum}

\clearpage


\clearpage

\setcounter{page}{1}
\renewcommand{\thefigure}{S\arabic{figure}}

\renewcommand{\thetable}{S\arabic{table}}
\setcounter{table}{0}

\noindent
{\LARGE Supplementary Information}

\section{Observations and Data Reduction}

We use spectroscopic data from two instruments in our analysis:
ground-based optical observations from the Ultraviolet Echelle
Spectrograph (UVES)\cite{Dekker2000} on ESO's Very Large Telescope
(UT-2), and space-based ultraviolet observations from the Space
Telescope Imaging Spectrograph (STIS) on-board the {\em Hubble Space
  Telescope}.

\subsection{UVES Data}

The UVES data presented here were taken in service mode on 2008
September 25 under the program 382.B-0556(A).  The observations
consist of 10 exposures of 2850 seconds each.  We used the $0\farcs7$
slit, feeding the light to dichroic \#2.  The grating central
wavelengths were 3,900 \AA\ in the blue and 7,600 \AA\ in the red.  No
binning was done on the chips.  The data were taken through thin
clouds and with seeing ranging from $\approx 1\farcs0$ to $1\farcs5$.

The data were reduced using the UVES pipeline provided by ESO.  In
order to achieve high signal-to-noise, we used the pixel-by-pixel flat
field correction, setting ``FFMETHOD=pixel'' in the reduction pipeline
and using optimal extraction of the data.  The individual exposures
were corrected for heliocentric motion and coadded, weighted by the
variance.  The signal-to-noise ratio ($SNR$) varies from $SNR\sim150$
per pixel in the far blue (e.g., near \ion{Ti}{2} 3,383) to $\sim300$
in some regions of the spectrum (e.g., the \ion{Ca}{1} and CH$^+$
transitions near 4,200 \AA).  Near the \LiI\ transition at
$\lambda\approx6,710$ \AA\ we measure $SNR\sim275$.

\subsubsection{UVES Line Spread Function:} The width of the UVES line
spread function (LSF) is important for deriving accurate fit
parameters for the blended isotopes of Li.  While ultimately the
column density of \LiI\ is not strongly changed by small changes in
the LSF width, the \bvalues\ are somewhat sensitive to the LSF width.
The UVES pipeline produces estimates of the width of the LSF on the
basis of observations of ThAr lamps taken for wavelength calibration.
For the blue CCD data, the ESO quality-control summaries available
on-line suggest the resolution for our set-up during this time is
$R\approx66,000$ or $\Delta v\approx4.54$ \kms, which is appropriate
for roughly the month surrounding our observations.  The ThAr lines in
our own calibration observations yield a value indistinguishable from
this.

On the red chip (REDL), notably in the vicinity of the \LiI\ lines
near 6,710 \AA, the mean FWHM of these lines from our calibration data
is 4.47 \kms\ for the 38 lines in the range $6,700 \, \mbox{\AA} \, <
\lambda < 6,720 \, \mbox{\AA}$ and 4.42 \kms\ for the 225 lines over
the broader range $6,650 \, \mbox{\AA} \, < \lambda < 6,770 \,
\mbox{\AA}$.  These correspond to a resolution $R\approx68,000$.

However, in the course of fitting the \LiI\ blend, adopting this
resolution seemed to underpredict the cores of the lines and provide
too much absorption immediately outside of the cores, suggesting the
adopted LSF may be too broad.  D. Welty kindly provided us with \NaI\
5,889/5,895 observations of the Sk~143 sight line taken with the CES
spectrograph on the ESO 3.6-m telescope with a resolution $\Delta v =
1.35$ \kms\ (FWHM) or $R\approx222,000$.  These data were presented in
Welty et al.\cite{Welty2006} We derived the best LSF breadth for the
current UVES red CCD data by comparing smoothed CES data with the UVES
observations of these absorption lines, using the $\chi^2$ as an
estimate of the goodness of fit.  The free parameters of the model in
this case were the breadth of the smoothing kernel (assumed Gaussian)
and the background offset of the CES observations (for which the
original background subtraction was derived by matching equivalent
widths of the \NaI\ lines with earlier UVES data\cite{Welty2006}).

The final result of this approach yields a best fit LSF breadth of
$\Delta v = 4.03\pm0.08$ \kms\ (FWHM) or resolution $R=74,300$, which
we adopt in fitting the \LiI\ blend.  The difference between our
derived LSF and the values from the ThAr emission lines is likely due
to the manner in which the light fills (or not) the slit during the
observations.  We caution that we assume in our fitting a single
Gaussian LSF; the true LSF may be more complex than this.  However, we
have tested the impact of the LSF on our measurement of both the
\lisevi\ column and isotopic ratio (see below), and find the results 
robust to even $5\sigma$ variations in the LSF breadth.  This LSF
breadth likely only applies to the data taken with the REDL chip.  LSF
changes with wavelength and detector are to be expected.  However, the
breadth of the LSF for the data taken with the blue side may be
smaller than the $4.55$ \kms\ predicted from the ThAr exposures.

\subsection{STIS Data}

We make use of archival STIS observations taken with the E140H and
E230H gratings using the $0.20 \arcsec \times 0.09\arcsec$ apertures.
These data were acquired under program 9383 (PI: K. Gordon) and have
been previously reported in the literature.\cite{Sofia2006} The
resolution of these data is $R \approx 114,000$ corresponding to
$\Delta v \approx 2.75$ \kms.  The data were processed with CALSTIS
v2.27 and otherwise reduced following the discussion in earlier
papers.\cite{Lehner2007} The $SNR$ of these data are limited by the
relatively low UV flux of this highly extincted star for high
resolution spectroscopy and ranges between 5 and 15 (per pixel) for
the lines of interest.

\section{Hydrogen column density along the Sk~143 sight line}

We derive the \HI\ column density toward Sk~143 by fitting the strong
\lya\ absorption seen in the {\em Hubble Space Telescope} (\hst)/STIS
E140H observations.  Following Lehner et al.,\cite{Lehner2008} we
fitted \lya\ with two components (Galactic and SMC) at fixed
velocities derived from the metal absorption lines and the \hi\ 21-cm
emission profile for the Galactic component (see below), which is in
agreement with the velocity from the absorption profiles.  Figure
\ref{fig:lyalpha} shows the STIS data for the region surrounding \lya\
(binned by 10 pixels for display purposes) along with our best fit
\HI\ profile and continuum.  For the SMC component we find $\log
N($\hi$) = 21.07 \pm 0.05$.  We tested the robustness of our fit by
undertaking several simulations wherein the velocity centroids of the
gas were changed by $\pm 5$ \kms, the placement of the continuum was
varied, and the continuum was modeled by a range of polynomials of
degree of 2 to 4.  The range of values derived in these simulations is
used as an estimate of the errors associated with these effects and is
included in our final uncertainty.  The column density of the Galactic
component in the fit is quite uncertain because it is much weaker (see
21-cm emission spectrum discussed below and in Figure
\ref{fig:radio}).  Its weakness is in fact a benefit for deriving an
accurate SMC $N($\hi$)$ as the damping wings are dominated by the SMC
component.

\setcounter{figure}{0}

\begin{figure}
\begin{center}\includegraphics[angle=0, width=6.in]{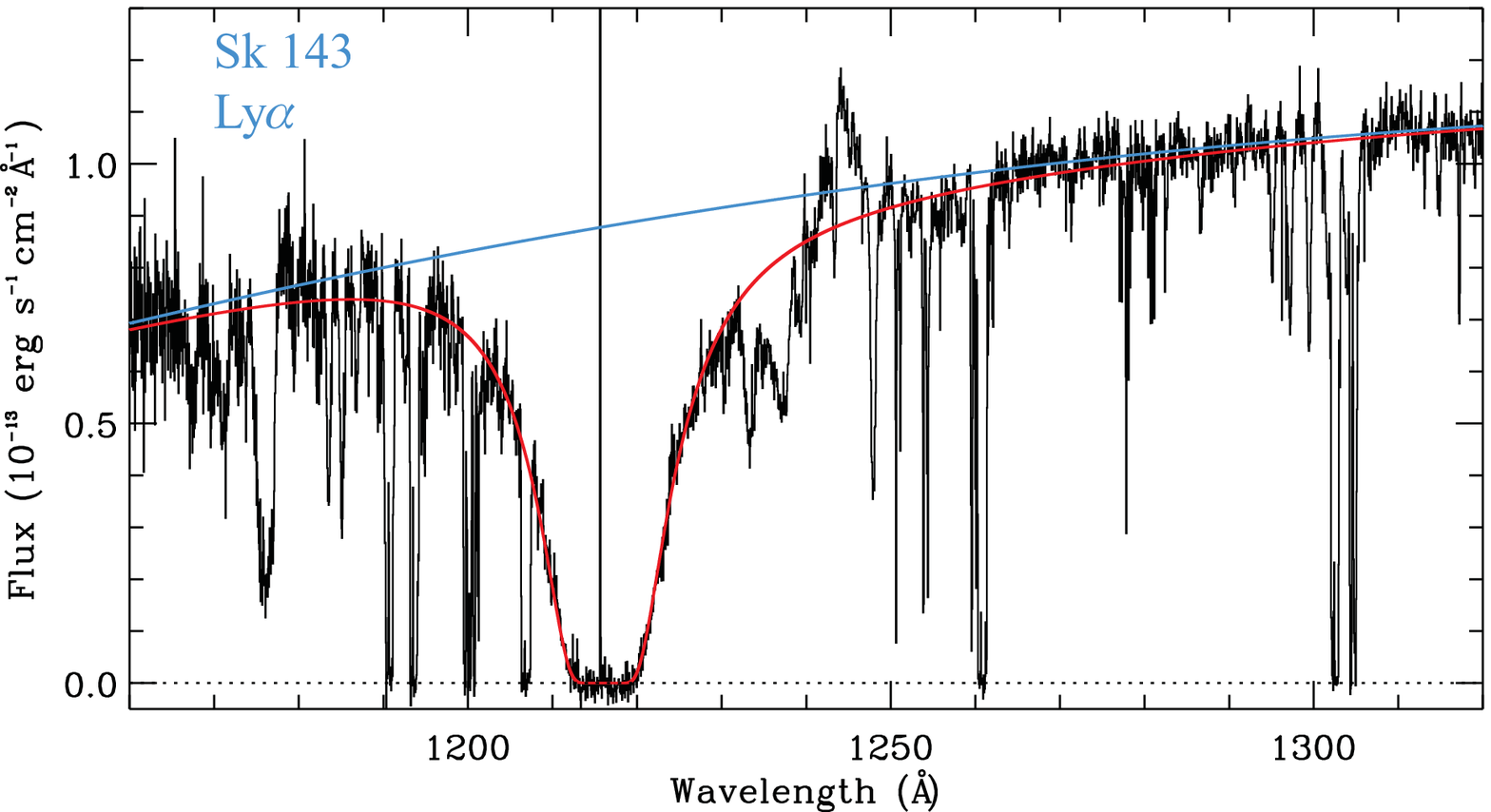}
\end{center}
\spacing{1.0}
\vspace{-0.75 cm}
\caption{\small {\bf Fit to interstellar \HI\ \lya\ absorption observed by
    \hst/STIS against the stellar continuum of Sk 143.} The blue line
  shows the adopted continuum, while the red line shows the continuum
  absorbed by an \HI\ column density of $\log
  N(\mbox{\HI})=21.07\pm0.05$. \label{fig:lyalpha} }
\end{figure}

\begin{figure}
\begin{center}
\includegraphics[angle=0, width=6.in]{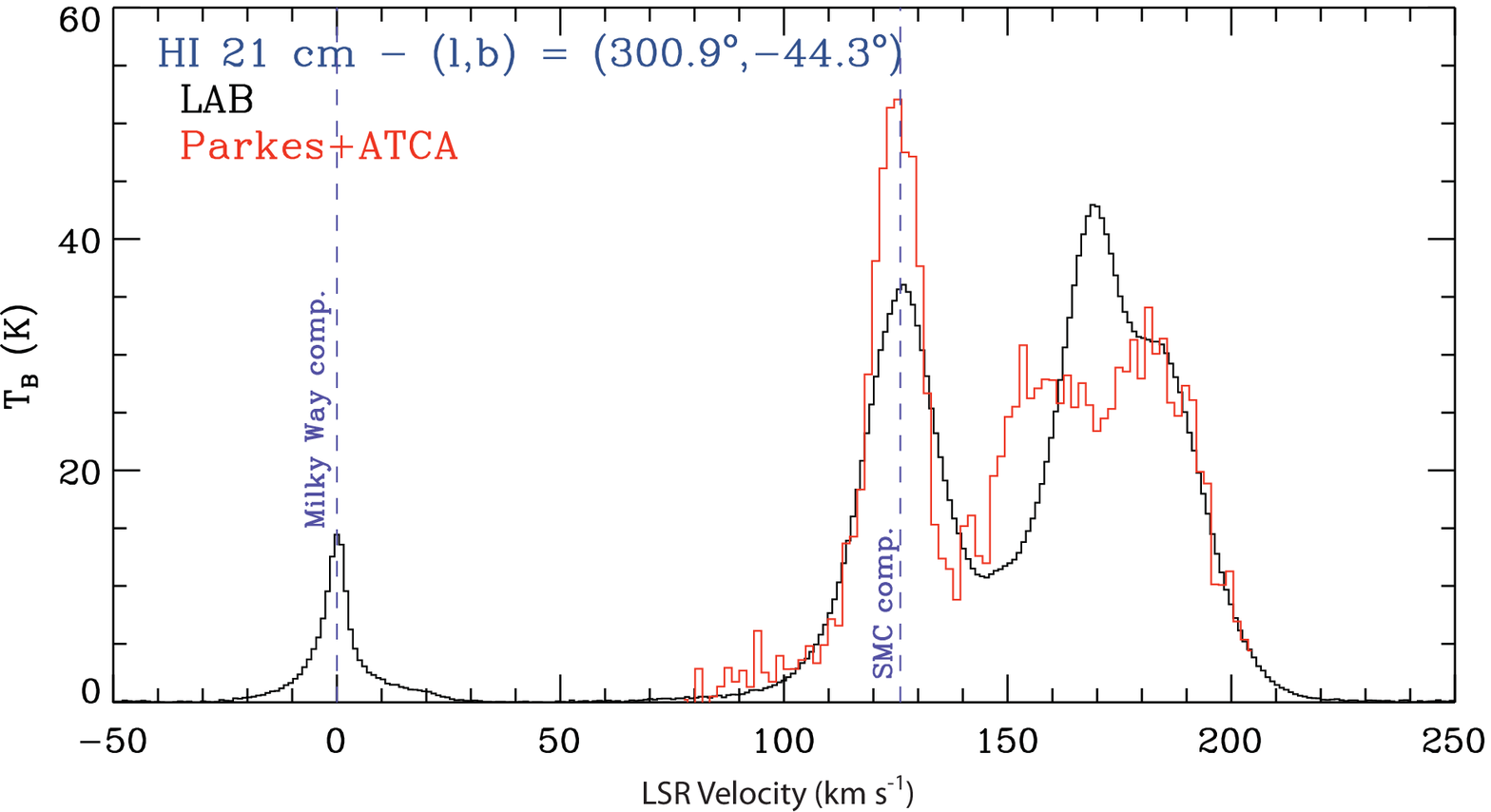}
\end{center}
\spacing{1.0}
\vspace{-0.75 cm}
\caption{\small {\bf \HI\ 21-cm emission line observations of the sight line
    to Sk 143.} The emission line data in black are taken from the
  Leiden/Argentine/Bonn (LAB) Survey\cite{Kalberla2005}, with a
  36\arcmin\ beam, while the data in red are taken from the
  Parkes/ATCA survey of the SMC\cite{Stanimirovic1999,
    Stanimirovic2004}, with a 1.6\arcmin\ beam.  The principal SMC
  component as seen in absorption is marked.\label{fig:radio} }
\end{figure}

Table \ref{tab:hicolumns} summarizes previous determinations of the
\HI\ column density in the SMC component toward Sk~143.  Previous
estimates relied on \hi\ 21-cm emission data or \lya\ absorption from
a low resolution and low signal-to-noise spectrum from the {\em
  International Ultraviolet Explorer}
(\iue).\cite{Gordon2003,Fitzpatrick1985,Bouchet1985} The poor quality
of the available \iue\ spectrum would seem to preclude a precise
measurement, and the first estimate using those data had a large
uncertainty of 40--50\%.\cite{Fitzpatrick1985,Bouchet1985} Most recent
studies of this sight line\cite{Cox2007,Welty2006,Gordon2003} have
adopted \HI\ estimates derived from the \iue -observed \lya\
absorption\cite{Fitzpatrick1985,Gordon2003}, or analyses of the \hi\
21-cm emission line
observations.\cite{Stanimirovic1999,Stanimirovic2004} The most recent
determinations using \iue\cite{Gordon2003} give high values of $\log
N($\hi$) \approx 21.6$, consistent with values derived by integrating
the full SMC profile of the \HI\ 21-cm emission seen in Figure
\ref{fig:radio}.  For comparison with absorption line data toward
Sk~143, however, the \hi\ 21-cm emission estimate is not always ideal,
as it includes contributions from gas behind the star.  Most of the
absorption toward Sk~143 is found at $\vlsr \sim 121$ \kms\ as seen in
Figures~\ref{fig:opticalstack} and \ref{fig:uvstack}. This component
is seen as well in 21-cm emission, but the 21-cm profile contains
significant amounts of gas to much larger velocities. Some of this is
seen in the absorption line observations, but with a smaller
contribution to the total column of undepleted elements (e.g., see the
total gas column traced by \ion{S}{2} absorption).  Integrating the
combined Parkes and ATCA \hi\
profile\cite{Stanimirovic1999,Stanimirovic2004} in
Figure \ref{fig:radio} over the velocity range containing the majority
of the optical/UV metal absorption, $\vlsr = 80$ to 146 \kms, we find
$\log N($\hi$) = 21.17 \pm 0.11$ (where the error is a combination of
statistical and ``beam'' errors).  Thus, the integration over this
smaller velocity range gives results consistent with our measurement
and another recent one of Welty \& Crowther\cite{Welty2010} at
$\la2\sigma$.  We note, though, that even though there is a kinematic
correspondence in the component at 121
\kms\ in the absorption and emission profiles, there is a possible
(unknown) uncertainty owing to the unknown depth of the star in the
SMC.  The moderate difference between our result and that of Welty \&
Crowther is due to the difference in the \HI\ column assigned to the
Milky Way, for which they derive values from 21-cm measurements while
we fit the \lya\ directly.


\setcounter{table}{0}
\begin{table*}[h]
{\small
  \caption{{\bf \HI\ column density estimates to Sk~143 using \lya}
    \label{tab:hicolumns}}
\begin{center}
\begin{tabular}{lll}
\hline
\hline
Reference & $\log N(\mbox{HI})_{\rm SMC}$ & Data \\
\hline
This Work                      
   & $21.07\pm0.05$ & \hst/STIS \\
Welty \& Crowther (2010)\cite{Welty2010}
   & 20.95          & \hst/STIS \\
Gordon et al. (2003)\cite{Gordon2003}
   & $21.60\pm0.05$ & \iue \\
Bouchet et al. (1985)\cite{Bouchet1985}
   & $21.40\pm0.17$ & \iue \\
Fitzpatrick (1985)\cite{Fitzpatrick1985}
   & 21.18$\pm0.18$ & \iue \\
\hline         
\end{tabular}
\end{center}
} 
\end{table*}

In what follows, we adopt our \HI\ column density measurement from the
\lya\ fitting.  With the \htwo\ column derived from \fuse\
data,\cite{Cartledge2005} the total hydrogen column along this sight
line is then $\log N({\rm H}) \equiv \log [N(\mbox{\HI})+2N({\rm
H_2})] = 21.46\pm0.04$.  The sight line has a high molecular fraction,
with $f({\rm H_2}) \equiv 2N({\rm H_2}) / N({\rm H}) = 0.6$.

\section{Metal column densities along the Sk~143 sight line}

Table \ref{tab:fullcolumns} gives our adopted final column density for
all of the atomic and ionic species detected toward Sk~143 along with
the methodology used to derive the columns.  We measure metal atom and
ion column densities along the Sk~143 sight line using a combination
of profile fitting and apparent optical depth integration methods,
depending on the quality of the data and the specifics of the
individual absorption lines.  We assume oscillator strengths from the
compilation of Morton.\cite{Morton2003} For each atomic or ionic
species, two column densities are given: the first applies to the
column density integrated over the entirety of SMC absorption, the
second gives the column density associated with the $+121$ \kms\
component within which the \LiI\ and other neutral species arise.  In
some cases (e.g., \LiI) these are the same as no significant
absorption exists outside of this cold, dense component.  We use the
column densities for this principal component at $\vlsr \approx
+121$ \kms\ in all calculations in the main text, since the \LiI
-bearing gas is all contained in this component.


\begin{table*}
\begin{center}
\begin{minipage}[b]{0.9\linewidth}
{\small
\caption{{\bf Column densities toward Sk 143}\label{tab:fullcolumns}}
\begin{tabular}{lccccc}
\hline
\hline
Species & $\log N(SMC)$ & $\log N(+121 \ \kms)$ & Method &
    $\abund{X}_\odot$ & [X/H]$_{\rm SMC}$\\
\hline
\HI     & $21.07\pm0.05$ & $21.07\pm0.05$ & 1 & $\equiv12.00$ & \nodata \\
H$_2$   & $20.93\pm0.09$ & $20.93\pm0.09$ & 2 & $\equiv12.00$ & \nodata \\
H$_{\rm total}$ & $21.46\pm0.04$ & $21.46\pm0.04$ & \nodata & $\equiv12.00$ & \nodata \\
\lisevi    & $10.29\pm0.02$ & $10.29\pm0.02$ & 3 & $3.23\pm0.05$ & $-0.55\pm0.16$ \\
\lisixi    & $9.41\pm0.15$  & $9.41\pm0.15$  & 3 & $2.14\pm0.05$ & $-0.38\pm0.21$ \\
\NaI    & $12.96\pm0.04$ & $12.96\pm0.04$   & 3 & $6.26\pm0.02$ & $>-2.76\pm0.06$ \\ 
\ion{Mg}{1} & $>13.11\pm0.06$ & $>13.11\pm0.06$ & 4 & $7.53\pm0.01$ & \nodata \\
\ion{Mg}{2} & $15.37\pm0.09$  & $15.37\pm0.09$  & 3 & $7.53\pm0.01$ & $-1.62\pm0.10$ \\
\ion{S}{1} & $14.15\pm0.08$    & $14.15\pm0.08$    & 3 & $7.14\pm0.02$ & \nodata \\
\ion{S}{2} & $\ga15.36\pm0.06$ & $\ga15.17\pm0.05$ & 3 & $7.14\pm0.02$ & $\ga-1.24\pm0.07$ \\
\KI        & $12.61\pm0.02$ & $12.61\pm0.02$ & 3 & $5.08\pm0.02$ & $>-1.93\pm0.04$ \\
%
%
\ion{Ca}{1} & $9.42\pm0.10$ & $9.42\pm0.10$   & 4 & $6.30\pm0.02$ & \nodata \\
\CaII       & $12.42\pm0.01$ & $11.68\pm0.01$ & 3 & $6.30\pm0.02$ & $>-3.34\pm0.04$ \\
\ion{Fe}{1} & $11.83\pm0.05$ & $11.83\pm0.05$ & 4 & $7.45\pm0.01$ & \nodata \\
\ion{Fe}{2} & $14.63\pm0.07$ & $14.15\pm0.10$ & 4 & $7.45\pm0.01$ & $-2.28\pm0.08$ \\
%
%
\ion{Ti}{2} & $11.88\pm0.03$ & $11.47\pm0.03$ & 4 & $4.92\pm0.03$ & $-2.50\pm0.05$ \\
\ion{Ni}{2} & $13.10^{+0.09}_{-0.11}$ & $12.97^{+0.08}_{-0.10}$ & 4 & $6.20\pm0.01$ & $-2.56\pm0.11$ \\
\ion{Zn}{2} & $\ga12.80\pm0.06$ & $\ga12.80\pm0.06$ & 4 & $4.60\pm0.03$ & $\ga-1.26\pm0.07$ \\
%
%
\hline         
\end{tabular}
\vspace{0.025in}\\
}
\spacing{1.0}
{\footnotesize Methods used for determining column densities: (1):
\lya\ Fitting; (2): Fitting by Cartledge et al.;\cite{Cartledge2005}
(3): Profile fitting; (4): $N_a(v)$ integration. The values of
[X/H]$_{\rm SMC}$ are gas-phase abundances, i.e., they reflect the
sub-solar SMC metallicity and the differential incorporation of
elements into the solid phase (dust grains).}
\end{minipage}
\end{center}
\end{table*}

\begin{figure}
\begin{center}
\includegraphics[angle=0, width=5.in]{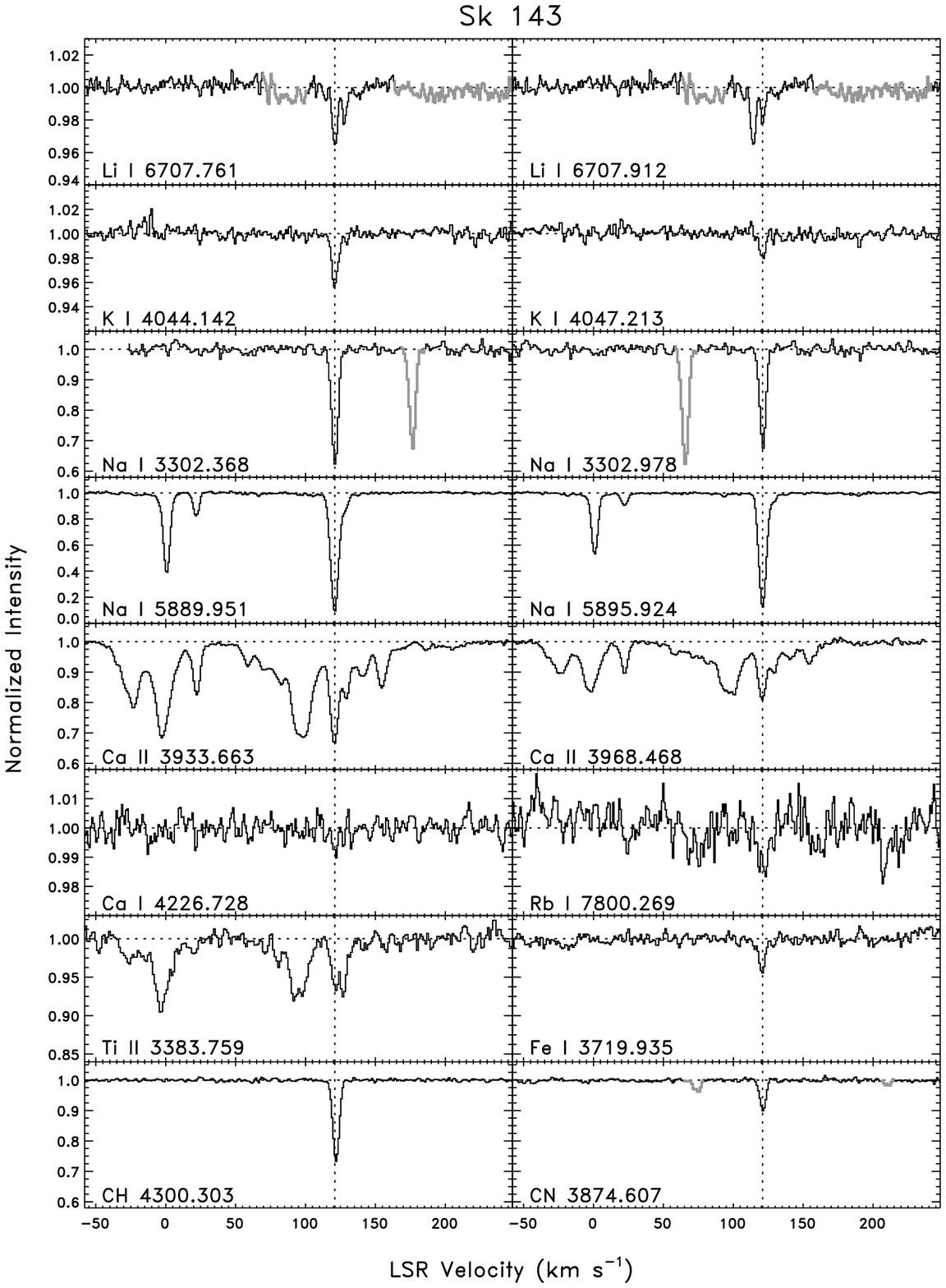}
\end{center}
\spacing{1.0}
\vspace{-0.75 cm}
\caption{\small {\bf Normalized interstellar absorption profiles for optical
    metal line transitions toward Sk 143 plotted versus LSR velocity.}
  These data are from the VLT+UVES observations of Sk 143.
  Contaminating absorption from other species are marked in light
  gray. \label{fig:opticalstack}}
\end{figure}
\begin{figure}
\begin{center}\includegraphics[angle=0, width=5.in]{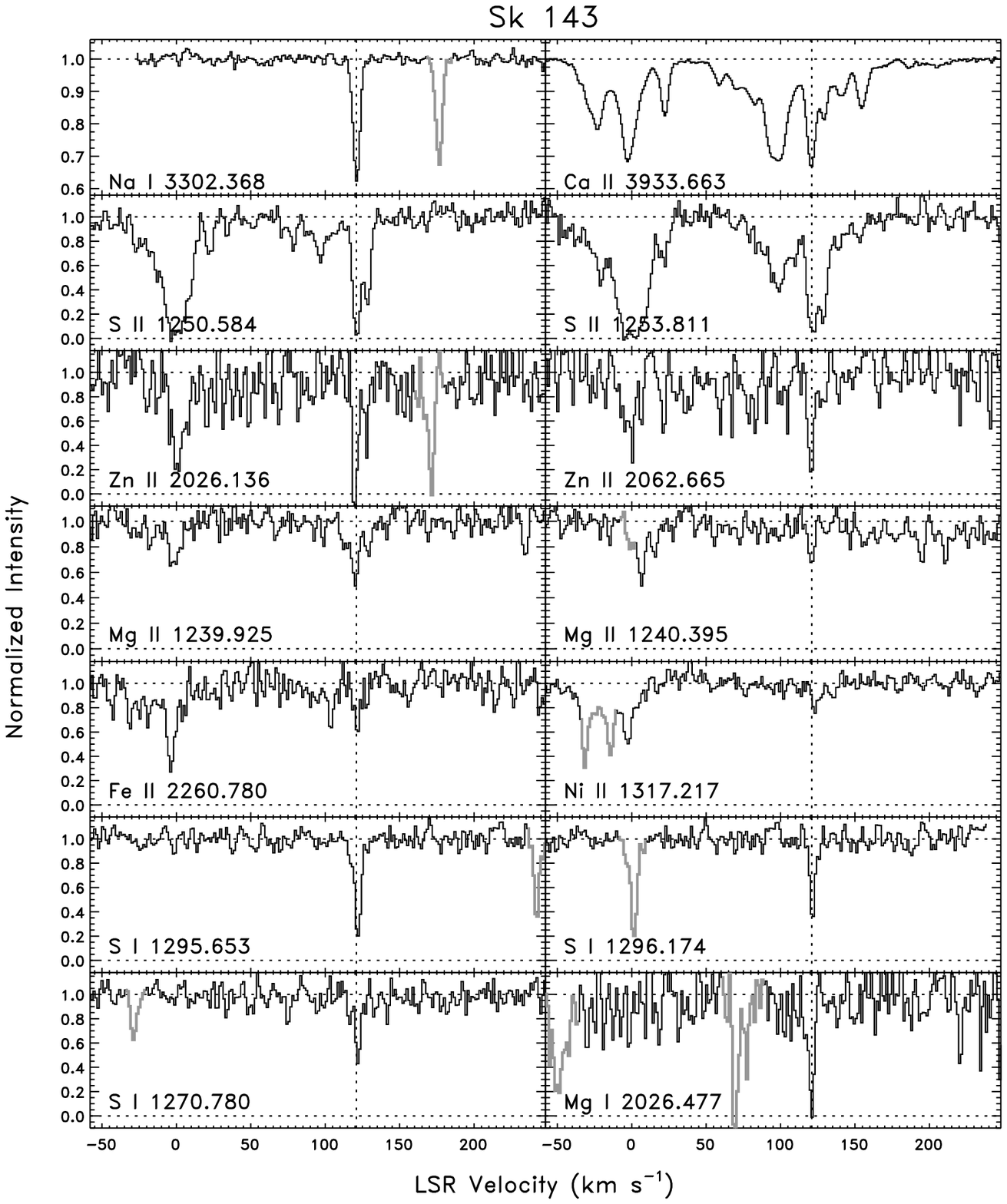}
\end{center}
\spacing{1.0}
\vspace{-0.75 cm}
\caption{\small {\bf Normalized interstellar absorption profiles for
    ultraviolet metal line transitions toward Sk 143 plotted versus
    LSR velocity.} These data are from the \hst/STIS E140H and E230H
  observations of Sk 143.  Contaminating absorption from other species
  are marked in light gray.  Optical observations from the previous
  figure are shown at the top for comparison.  \label{fig:uvstack}}
\end{figure}

Our component fitting column densities are derived using the software
package VPFIT\cite{Webb1987} made available by R. Carswell
(http://www.ast.cam.ac.uk/~rfc/vpfit.html).  VPFIT minimizes the
$\chi^2$ goodness of fit parameter using a Gauss-Newton type system
for parameter updates.\cite{Murphy2002} The free parameters of the fit
for each assumed interstellar ``component'' are the central velocity,
Doppler parameter ($b$-value), and column density of the component.
We fit each species independently, with no assumptions about
$b$-values or central velocities from other absorbing species (as
opposed to assuming the $b$-values are linked through a common
temperature and non-thermal velocity).  This has the advantage that
additional unsuspected components should not adversely affect the
results, so long as the lines are not strongly saturated.  We fit
multiplets simultaneously where available.  We did not, however,
simultaneously fit the strongest doublets of \NaI\ and \KI\ with the
weakest doublets available (see Figures \ref{fig:opticalstack}
and \ref{fig:uvstack} for the observed lines).  The column densities
in Table \ref{tab:fullcolumns} are derived based on the weaker
doublets of these species due to the extreme saturation of the
strongest lines.

Our fitting of all of the neutral species suggests that the SMC
absorption seen in weak lines of neutral species (i.e., not including
the \NaI\ D lines) are well described by a single component.  We see
no compelling evidence for additional components that contribute
significantly to the column densities of the neutral species.  The
high resolution CES data presented by Welty et al.\cite{Welty2006} are
consistent with a single dominant component with only very weak
additional components at higher relative velocities.  We note that the
presence of additional unknown components could affect the
interpretation of the Li isotopic ratio if they had strong variations
in the ratio between components or if there were moderate strength
components present with separations matching the isotope shift.  For
derivation of total column densities of \lisev\ (and largely even
\lisix), the presence of multiple components is not particularly
important.  In general our profile fits to these neutral species gave
\bvalues\ of order 0.7 \kms, consistent with previous studies of this
sight line.\cite{Welty2006,Cox2007} We note also that the presence of
additional unidentified components is unlikely unless they have a
separation less than this 0.7 \kms.

Our derived column densities of \NaI, \KI, and \CaII\ vary somewhat
from these earlier works\cite{Welty2006,Cox2007} (with our columns
being higher) because we have generally studied weaker lines, fully
fit multiple lines, or in the case of \CaII\ used a careful treatment
of the continuum for the weak line, which lies on the stellar
H$\epsilon$ absorption requiring special techniques for the continuum
assessment.

Our fitting assumed Gaussian LSFs with breadths derived as described
above for the UVES observations of \LiI.  For the blue CCD data we had
no higher resolution comparison observations, so we adopted the LSF
breadth suggested by the calibration lamp data.  There was no obvious
indication that this LSF is too broad.  For fitting the STIS E140H
observations of \ion{Mg}{2}, we assumed a breadth of 2.75 \kms\
(FWHM).  The STIS LSF is not strictly a Gaussian.  In our case we were
only interested in separating the SMC \ion{Mg}{2} absorption from the
contaminating Milky Way absorption. For these purposes and with the
strength of the lines, the assumption of a Gaussian LSF will not
adversely affect the results.

We have used profile fitting to derive column densities for neutral
species with small $b$-values in case unresolved saturation is present
or where blending is a concern.  For most of the ionic species we
adopt column densities derived through an integration of the apparent
column density, $N_a(v)$, profiles,\cite{Savage1991} largely following
methodologies in our earlier works.\cite{Howk2003} Saturation can be
identified by comparing the \nav\ profiles from two lines of an
absorbing species with significantly different oscillator strengths
(\fvalues).  The entries in Table \ref{tab:fullcolumns} listed as
lower limits are those for which we have evidence for unresolved
saturation.

Table \ref{tab:fullcolumns} also gives the solar system
abundances\cite{Asplund2009} for each element and the relative
gas-phase abundance of the element in the ISM of the SMC, [X/H]$_{\rm
SMC}$.  The latter quantity reflects the base subsolar metallicity of
the present-day SMC, [Fe/H]$_{\rm SMC} = -0.59\pm0.06$ derived from a
straight mean of observations of early type stars,\cite{Trundle2007,
Bouret2003, Korn2000, Luck1998} and the effects of differential dust
depletion of the elements.\cite{Jenkins2009} Where two ionization
states are measured, we only give the gas-phase values for the most
abundant ion.  Inequalities either reflect the non-dominant nature of
the species (e.g., for the neutrals) or the probable presence of
unresolved saturation.  The abundances listed for the Li isotopes are
drawn from the main text (i.e., adopting the abundances derived from
scaling the \LiI /\KI\ ratio).

Assessing the effects of dust depletion on the gas along this sight
line is somewhat complicated by the lack of a quality measurement of
an undepleted metal species.  The [X/H] values in Table
\ref{tab:fullcolumns} measure the effects of the sub-solar metallicity
of the SMC and its non-solar relative abundance pattern in addition to
dust depletion effects.  Welty \& Crowther\cite{Welty2010} have
recently estimated the intrinsic Ti abundance of the SMC based on
early-type stars, deriving an estimate for the present day abundance
of \abund{Ti}$_{\rm SMC} \approx 4.28$.  Our measurements of the
gas-phase \ion{Ti}{2} then suggest Ti is depleted by $-1.86$ dex
relative to the intrinsic SMC abundance; sight lines in the Milky Way
have depletions $-2$ to $-3$ for similar $E(B-V)$ or CN columns, but
it is not dissimilar from sight lines in the Galactic disk with modest
depletions.\cite{Welty2010} Defining the reference abundance can be
difficult for assessing the depletion of an element.  Another approach
to understanding the level of elemental incorporation into grains is
to measure the differential depletion between two metals.  The least
depleted species with a good measure of the total column density in
the table is that of Mg, for which we have a good measure of the
dominant ion \ion{Mg}{2}.  Comparing strongly-depleted metals to Mg in
the SMC gas toward Sk~143 gives [Fe/Mg]$_{\rm SMC} = -0.66\pm0.13$,
[Ni/Mg]$_{\rm SMC} = -0.94\pm0.15$, and [Ti/Mg]$_{\rm SMC} =
-0.88\pm0.11$.  The uncertainties are large given the low SNR of the
STIS data for deriving abundances, so comparing these data to Milky
Way sight lines does not lead to great insight.  These are consistent
with relative gas-phase abundances seen in Milky Way disk gas for
sight lines with low to modest depletion (at least for the
disk).\cite{Jenkins2009,Cartledge2006,Savage1996} Measurements of Li
in the Milky Way typically probe sight lines or clouds with stronger
depletions than these values; the Li depletion effects should not as
strong in this SMC gas as in the gas typically probed in the Milky
Way.



\section{Lithium abundance estimates}

Several estimates for the gas-phase \lisev\ abundance are given in the
text, which are summarized in Table \ref{tab:finalresults}.
Estimating the gas-phase \lisev\ abundance from measurements of
$N(\mbox{\lisevi})/N({\rm H})$ requires the application of an
ionization correction to account for the unseen \lisev, which is
mostly in its singly ionized form in the ISM.  Assuming ionization
rate balance and only atomic processes, $N({\rm Li})/N(\mbox{\LiI})
\approx N(\mbox{\ion{Li}{2}})/N(\mbox{\LiI}) = \Gamma(\LiI) [n_e
\alpha(\mbox{\ion{Li}{2}},T)]^{-1}$, where $\Gamma(\LiI)$ is the
photoionization rate and $\alpha(\mbox{\ion{Li}{2}},T)$ the
recombination coefficient.  (We generically refer to \LiI\ here, which
applies to either isotope or their sum.)  At first glance it would
seem that knowledge of $\Gamma$ and $n_e$ are crucial to estimating
the total abundance.  However, similar equations can be written for
observed adjacent ionization states, notably for
$N(\mbox{\ion{Ca}{2}})/N(\mbox{\ion{Ca}{1}})$ and
$N(\mbox{\ion{Fe}{2}})/N(\mbox{\ion{Fe}{1}})$.  Substituting $n_e$
derived from these ratios into the expression for Li gives, for
example,
\begin{displaymath}
  \frac{N(\mbox{\ion{Li}{2}})}{N(\mbox{\LiI})} = 
  \frac{N(\mbox{\ion{Fe}{2}})}{N(\mbox{\ion{Fe}{1}})}
  \frac{\Gamma(\LiI)/\Gamma(\mbox{\ion{Fe}{1}})}
  {\alpha(\mbox{\ion{Li}{2}},T)/\alpha(\mbox{\ion{Fe}{2}},T)} .
\end{displaymath}
While the individual photoionization rates depend on the intensity of
the interstellar radiation field, to first order the ratio of
photoionization rates $\Gamma(\LiI) / \Gamma(\mbox{\ion{Fe}{1}})$
does not.  What is crucial in this case is the shape of the radiation
field over the range of energies encompassed by the ionization edges
of these neutrals.  The grossly similar temperature sensitivities for
the recombination coefficients mitigates the effect of this unknown,
as well.  The total abundance of Li is then $\log
[N(\mbox{\lisev})/N({\rm H})] = \log [N(\mbox{\lisevi})/N({\rm H})] -
\log x(\LiI)$, where the term $- \log x(\LiI)$ is the ionization
correction factor and $x(\LiI) \equiv N(\LiI)/N({\rm Li})$ is the
ionization fraction of Li in neutral form. 


\begin{table*}
\begin{center}
\begin{minipage}[b]{0.65\linewidth}
{\small
\caption{{\bf Summary of SMC lithium abundances}\label{tab:finalresults}}
\begin{tabular}{lcl}
\hline
\hline
Quantity   & Value   & Methodology \\
\hline
\abund{\lisev}$_{\rm SMC}$ & $2.68\pm0.16$ & Scaled from \lisevi /\KI \\
\abund{\lisev}$_{\rm SMC}$ & $2.38\pm0.17$ & Scaled from \lisevi /\SI \\
\abund{\lisev}$_{\rm SMC}$ & $3.01\pm0.12$ & 
                     Ionization correction from \ion{Ca}{2}/\ion{Ca}{1} \\
\abund{\lisev}$_{\rm SMC}$ & $2.79\pm0.11$ & 
                     Ionization correction from \ion{Fe}{2}/\ion{Fe}{1} \\
$[\mbox{\lisev/K}]_{\rm SMC}$ & $+0.04\pm0.10$ & 
                     Differential ionization correction \\
$[\mbox{\lisev/S}]_{\rm SMC}$ & $-0.26\pm0.11$ & 
                     Differential ionization correction \\
\hline         
\end{tabular}
\vspace{0.025in}\\
}
\spacing{1.0}
{\footnotesize All absolute and relative abundances of SMC Li derived using several
different methods are given.  The recommended values are those derived
by comparison with K, i.e., \abund{\lisev}$_{\rm SMC} = 2.68\pm0.16$
and $[\mbox{\lisev/K}]_{\rm SMC} = +0.04\pm0.10$.  The former relies
on the application of several scale factors, including a differential
ionization correction, the metallicity of the SMC, and an estimated
K/Fe abundance.  The latter relies on a differential ionization
correction, but one that is well founded in the observed Galactic
relationship between \LiI\ and \KI.}
\end{minipage}
\end{center}
\end{table*}


As discussed in the main text, we derive ionization corrections for
the Li/H comparison on the basis of the observed ratios
$N(\mbox{\ion{Ca}{2}})/N(\mbox{\ion{Ca}{1}})$ and
$N(\mbox{\ion{Fe}{2}})/N(\mbox{\ion{Fe}{1}})$.  These give ionization
corrections of $-\log x(\LiI) = +1.96\pm0.10$ and $+2.18\pm0.12$ from
the Ca and Fe ions, respectively, assuming standard relative
ionization rates with specific interstellar radiation
fields.\cite{Welty2003,Pequignot1986} These are derived using column
densities from the $\vlsr \approx +121$ \kms\ cloud only.  The ratio
of \lisevi\ to total hydrogen in the SMC gas along this sight line is
$\log [N(\mbox{\lisevi})/N({\rm H})] = -11.17\pm0.04$, giving $\log
[N(\mbox{\lisev})/N({\rm H})] = -9.21\pm0.11$ and $-8.99\pm0.13$.
While we also have data covering adjacent ions of S and Mg, we have
only limits for one ionization state of each of these elements that do
not produce useful limits on the ionization corrections.

The discrepancies between these calculations are in part due to errors
in the calculated photoionization rates and recombination
coefficients.  However, it is also likely that the neutral and
singly-ionized Ca and Fe, which have different ionization potentials,
photoionization cross sections, and recombination coefficients, trace
slightly different conditions within the cloud.  The conditions traced
by these species will also be different than those traced by \LiI\ and
\ion{Li}{2}, limiting the precision of this approach.  In fact,
electron densities derived from various tracers can vary, with
\ion{Ca}{1} typically predicting significantly larger $n_e$ than
most.\cite{Welty2003} In the SMC cloud probed here, however, the
\ion{Ca}{1} column seems to be unusually low, as indicated by the
\ion{Ca}{1} to \ion{K}{1} ratio, for example, compared with Milky Way
sight lines. The integrated
$N(\mbox{\ion{Ca}{2}})/N(\mbox{\ion{Ca}{1}})$ is higher than all
measured values in the compilation of Welty et al.\cite{Welty2003}
(although some sight lines with nondetections of
\ion{Ca}{1} could have higher values), although the ratio in the 
$\vlsr \approx +121$ \kms\ component is near the mean of those
measurements.  

Using \LiI /\KI\ as a tracer of the total interstellar Li/K abundance
follows a similar approach:
\begin{displaymath}
  \frac{N(\mbox{\ion{Li}{2}})}{N(\mbox{\ion{K}{2}})} = 
  \frac{N(\mbox{\ion{Li}{1}})}{N(\mbox{\ion{K}{1}})}
  \frac{\Gamma(\LiI)/\Gamma(\mbox{\ion{K}{1}})}
  {\alpha(\mbox{\ion{Li}{2}},T)/\alpha(\mbox{\ion{K}{2}},T)} .
\end{displaymath}
In this case the ratio of relative ionization rates to recombination
coefficients gives an ionization correction of $+0.54\pm0.08$
(following Steigman\cite{Steigman1996} in adopting an error based on
the range of recombination coefficients derived from a wide range of
temperatures).  The \LiI/\KI\ ratio is powerful given the very nearly
linear relationship between the column densities of these species in
the Milky Way.\cite{Knauth2003, Welty2001} We assume the same physics
applies in the SMC, and thus this ratio should be similarly useful
there. This implies Li and K have very similar dust
depletion\cite{Knauth2003, White1986} and ionization conditions that
track each other well.  Similarly, the \KI /\SI\ ratio is a power law
with an index somewhat greater than unity.\cite{Welty2003} The tight
correlation with larger power law slope likely implies the two
neutrals trace similar physical conditions, but with different
depletion characteristics.  We make the assumption that \LiI\ also
traces physical conditions similar to those of \SI -bearing clouds.
The ionization correction relating \LiI /\SI\ to Li/S is
$-0.32\pm0.08$ dex.  We do not use the other neutral tracers for such
a derivation of the Li to metal nuclei ratio because of the weaker
correlations and evidence for differeng physical conditions between
\LiI\ and the remaining neutrals.

Transforming [\lisev /K]$_{\rm SMC}$ to an estimate of
$\abund{\lisev}_{\rm SMC}$ involves a number of scale factors.  Our
adopted mean metallicity of the SMC ([Fe/H]$_{\rm SMC}$) is described
above.  One other scale factor is the abundance of K/Fe with respect
to the solar system.  The K/Fe ratio is not measured directly in the
SMC.  However, K has its origins in explosive oxygen burning in high
mass stars and tends to behave nucleosynthetically like an $\alpha$
element.\cite{Samland1998} Studies of early-type
stars,\cite{Hunter2009,Dufton2005,Korn2000} A
supergiants,\cite{Venn1999} and \ion{H}{2} regions\cite{Peimbert2000}
suggest $[{\rm S,Si/Fe}] \approx 0$ in the present-day SMC. These
studies can have significant uncertainties, and there is some
dispersion in the results.  We adopt a value $[{\rm K/Fe}]_{\rm SMC} =
+0.0\pm0.1$ in our analysis (uncertainty from dispersion in the
$\alpha$ to Fe dispersion in the SMC).  We note that this is at odds
with what is seen in the Milky Way.  The mean of two samples of Milky
Way stars within $\pm0.1$ dex of the SMC
metallicity,\cite{Zhang2006,Takeda2002} $[{\rm K/Fe}]_{\rm SMC} =
+0.27\pm0.09$ (standard deviation).  The discrepancy is due to the
significantly different star formation and enrichment history of the
SMC compared with the Milky Way.\cite{Tolstoy2009}

Our \lisev\ abundance and \lisev /K ratio is significantly higher than
that implied\cite{Steigman1996} by the upper
limits\cite{Baade1991,Baade1988,Sahu1988} or claimed
detections\cite{Vidal-Madjar1987} toward SN1987A in the Large
Magellanic Cloud (LMC). With an abundance roughly twice that of the
SMC, the LMC should have a \lisev\ abundance at least as high as that
of the SMC.  The values reported for this sight line vary
significantly, and we view them with significant skepticism in
hindsight.  Aside from the discrepancies in the \LiI\ measurements,
there are worries about saturation corrections for the \KI\ reference.
Furthermore, the sight line to SN1987A probes a significantly lower
column density, more highly ionized portion of the ISM.  This sight
line does not favor the formation of atomic species such as \LiI,
making the ionization effects even more severe.  Future observations
of \LiI\ absorption within the LMC should provide more clarity to the
chemical evolution of Li in the Clouds.


\section{Lithium Isotopic Ratio}

The Li isotope ratio can provide some constraints on the production of
Li and any non-standard contributions to Li since \lisix\ is not
expected to be produced in significant amounts in
BBN.\cite{Pospelov2010} Our estimate of the \liratio\ relies on
simultaneously fitting the hyperfine structure of the \lisevi\
and \lisixi\ absorption.  We adopt wavelengths for the hyperfine
levels from Sansonetti et al.\cite{Sansonetti1995} and \fvalues\
summarized in Welty et al.\cite{Welty1994} We assume the ISM
absorption from the neutral species is well-characterized by a single
component or cloud and that both isotopes have a common \bvalue.  The
latter assumption is only strictly correct if the velocity dispersion
is dominated by non-thermal motions (which is likely), although it has
little practical effect given the weakness of the \LiI\ absorption.
We fit a stellar continuum to the data before the fitting process.
The continuum in the region around \LiI\ was well fit with a
first-order Legendre polynomial.  We assessed higher order fits using
an $F$-test, but they did not provide a statistically significant
improvement in the goodness-of-fit parameter over the region used to
derive the continuum.  In the profile fitting process, we allow the
parameters of the continuum fit to vary during the minimization
process.  This allows VPFIT both to esimate the continuum fit
objectively and to include an estimate of the uncertainties caused by
continuum fitting into the final error budget.

The column densities reported in Table \ref{tab:fullcolumns} are
derived from these fits.  The best fit Doppler parameter is $b =
0.8\pm0.5$ \kms.  Because the absorption is so weak, the \bvalue\ has
a minor effect on the quality of the column density determination (as
would multiple components).  The fits were made with the UVES LSF as
discussed above.  The resolution we derive for the red side data are
higher than typically advertised or adopted for the instrument.  Our
error budget includes a contribution derived by refitting the data
with LSF breadths varied by $\pm1\sigma$, but the resulting
differences in the central value are minimal.  Adopting a
significantly broader FWHM for the LSF (e.g., $\Delta v \approx
4.45$ \kms\ as suggested by the ThAr lines) has very little effect on
the central value or uncertainties, although the $\chi^2$
goodness-of-fit parameter is larger.  In this case, the profiles
appear visibly broader than the data, with the peaks of the models
underpredicting the observed absorption in the core and the wings
overpredicting the observed absorption in the edges of the profiles.
We have tested the impact of adding additional components, but VPFIT
rejects them as unnecessary as they do not significantly improve the
fit.

It is important to point out that the potential presence of \lisixi\
absorption is not simply a function of the manner in which we have
proceeded with the profile fitting.  The \lisixi\ contribution is
hinted at from simple equivalent width measurements of the \lisevi\
doublet.  Using our initial continuum (i.e., not the final one adopted
above through the formal fitting process), the weak and strong lines
of \lisevi\ give equivalent widths $W_\lambda = 3.77\pm0.23$ m\AA\ and
$2.40\pm0.22$ m\AA.  (We caution both the errors and equivalent widths
derived here may not be equivalent to those reported through the
profile fitting analysis due to the different manner in which the
continuum placement uncertainties are included and the crude nature of
the separation of the strong and weak lines in the direct integration
of equivalent widths.)  Assuming there is no saturation, a safe
assumption given the weakness of the lines, these lines should have
equivalent widths in a 2:1 ratio.  The observed ratio of 1.57:1 is
lower due to the blended contribution of \lisixi\ to the weak member
of the \lisevi\ doublet.  A simple estimate of the \lisixi\ column
density can be derived by subtracting half of the \lisevi\ strong line
equivalent width from the weak line value.  This yields $\log
N(\lisixi) \approx 9.41\pm0.20$, completely consistent with the value
derived from the profile fitting.

The derived isotopic ratio for the SMC gas toward Sk~143 is
$\liratio_{\rm SMC} = 0.13\pm0.05$, which implies a limit
$\liratio_{\rm SMC} < 0.28$ ($3\sigma$).  The limit itself does not
constrain the ratio in an interesting way.  It is, for example,
consistent with the Solar System value, $(\lisix /\lisev)_\odot =
0.0787\pm0.0004$,\cite{Rosman1998} and it is very similar to a number
of measurements in Milky Way gas,\cite{Kawanomoto2009, Knauth2003}
although there are clouds and sight lines with large variations from
this value.\cite{Knauth2000} Kawanomoto et al.\cite{Kawanomoto2009}
have recently derived a mean for the Milky Way ISM of $\langle \lisix
/\lisev \rangle_{\rm MW} = 0.13\pm0.04$.  The only known source of
post-Big Bang \lisix\ is production via cosmic ray interactions with
interstellar medium (ISM) particles, either through spallation or
$\alpha+\alpha$ fusion.\cite{Reeves1970,Meneguzzi1971,Steigman1992}
For standard energy distributions of Galactic cosmic rays, these
isotopes are produced in a ratio $\liratio \approx 0.67$, while
non-standard assumptions can push this ratio to 0.5 or a bit
lower.\cite{Indriolo2009} For the standard production ratio, our
central value implies that $(19\pm8)\%$ of the \lisev\ in the ISM of
the SMC has been produced via cosmic rays, with a limit of $<42\%$
($3\sigma$).

Our data show \lisixi\ at $<3\sigma$ significance, and the reality of
the \lisixi\ at the levels of our best fit value should rightfully be
viewed with caution. 
%
%
Sk~143 is bright enough that follow-up observations should be able to
significantly reduce the statistical uncertainties without much
difficulty.  We have not pushed the limits of UVES with the present
observations.  At the same time, measurement of $\abund{\lisix}$ in
sub-solar metallicity environments are potentially incredibly
important.  A significant abundance of \lisix\ in low metallicity
environments could come from pre-galactic cosmic rays accelerated via
the collapse of large scale structure (producing Li via
$\alpha+\alpha$ fusion)\cite{Suzuki2002} or even from the effects of
non-Standard Model particles in the early universe that affect
BBN.\cite{Pospelov2010} Future observations of the gas along this and
other SMC or LMC sight lines may be the best way to reliably constrain
the Li isotopic ratio at sub-solar metallicities.

\section{Notes on Chemical Evolution Models}

We compare our results in Figures \ref{fig:liabundance}
and \ref{fig:li2metal} with those of recent chemical evolution models
of Prantzos,\cite{Prantzos2012} although there are other models in the
literature.\cite{Romano2003, Casuso2000, DAntona1991, Mathews1990}
These models account for post-BBN contributions to the total \lisev\
abundance from massive stars through neutrino nucleosynthesis in core
collapse supernovae\cite{Woosley1990} and from low mass stars as red
giant or asymptotic giant branch stars or novae.\cite{Prantzos2012}
The \lisev\ yields from these sources are poorly known.  Both \lisev\
and \lisix\ are synthesized directly in the ISM through cosmic ray
nucleosynthesis.

The Prantzos models have worked around the uncertainties in the
stellar yields by assuming the low-mass star contributions are
sufficient to reproduce the solar system (meteoritic) abundances.  The
relative contributions of the various sources as a function of
metallicity follow the relative importance of core collapse (high
mass) and type Ia (low mass) supernovae (SNe).  The cosmic ray model
adopted follows from an analysis of the the energy input from core
collapse SNe of massive stars that have suffered mass loss through
stellar winds.\cite{Prantzos2012} In assessing the yields, no credence
is given to the Li abundances of stars in the Milky Way.  The only
data points matched by the models are the primordial and meteoritic
abundances of Li.

While the uncertainties in the yields are significant, these models
are illustrative of the shape of the post-BBN evolution of cosmic Li.
However, the primordial and solar system abundances adopted by
Prantzos are different than those adopted in our work.  Thus, for
comparison with our measurements, we have had to make adjustments to
the model results.  We have done this by adopting the Cyburt et
al.\cite{Cyburt2008} primordial abundance, assuming the contributions
from core-collapse SNe and cosmic rays are as given in Prantzos' work,
and independently scaling his adopted low-mass star production rates
to match our adopted solar system abundance.\cite{Asplund2009}

\vspace{0.3in}

\hrule

\end{document}